\documentclass[11pt]{article}             
\title{System with classical and quantum subsystems in tomographic
probability representation}
\author{V.~N.~Chernega$^{\dag}$, V.~I.~Man'ko$^{\dag}$
        \\$^{\dag}$ P.N.~Lebedev Physical Institute, Russian Academy of Sciences\\
        Leninskii Prospect, 53, Moscow 119991, Russia \\
        \\Emails: vchernega@gmail.com, manko@sci.lebedev.ru}

\begin{document}

\maketitle

\begin{abstract}
Description of system containing classical and quantum subsystems by
means of tomographic probability distributions is considered.
Evolution equation of the system states is studied.
\end{abstract}

\section{Introduction}
States of a classical particle moving in an environment are
described by a probability density $f(q,p)$ on the particle
phase-space where position $q$ and momentum $p$ fluctuate. Pure
states of a quantum particle are associated with complex wave
function $\psi(x)$ \cite{Sch26}. For the quantum particle in an
environment the states are identified with the Hermitian nonnegative
density operator $\hat\rho$ (or its density matrix, e.g., in
position representation $\rho(x,x')$ which is the complex function
of two real variables \cite{Landau27,vonNeumann27}. Recently, the
tomographic probability representation of both quantum
\cite{ManciniPhysLet96,Ibort2009} and classical
\cite{Olga1997,MendesVIMPhysicaD} states was introduced. In this
representation the classical particle states and the quantum
particle states are identified with tomographic probability
distribution $w(X,\mu,\nu)$ (called symplectic tomogram) or
$w(X,\theta)$ (called optical tomogram) where the random variable
$X$ is the particle position measured in the corresponding reference
frame of the particle phase-space. The reference frame is labeled by
the two real parameters $\mu$ and $\nu$ for symplectic tomogram and
by the rotation angle $\theta$ for the optical tomogram. The
rotation angle $\theta$ of the reference frame axes in the
phase-space is called the local oscillator phase \cite{Raymer} in
the cases where the photon quantum states are considered and their
Wigner functions are reconstructed \cite{LvovskiRaymer} using
experimental results providing the optical tomogram $w(X,\theta)$
(see, e.g. \cite{SolimenoPorzio}). In \cite{Olga2004} it was pointed
out that since the classical and quantum particle states are
identified with the same tomographic probability distribution, e.g.
symplectic tomogram there exists a possibility to construct the
quantum and classical mechanics in a framework of a unified scheme
namely using the tomographic probability representation. This idea
to suggest a scheme where both classical and quantum mechanics are
unified was discussed in the literature before see, e.g.
\cite{Peres199?}. Recently Elze with collaborators \cite{Elze1}
suggested the description of both classical and quantum linear
dynamics using phase-space representation of the particle and path
integral formalism. The problems of such unified construction of the
mechanics which combines both classical and quantum states are
related to difference of the states description in classical and
quantum domains. The Wigner function $W(q,p)$ \cite{Wig32} is
similar to classical probability density $f(q,p)$ but nevertheless
it is not nonnegative probability distribution of measurable
particle position and momentum. The aim of this work is to use the
particle state description both in classical and quantum domains by
the tomographic probability distribution and to suggest the
description of the states of systems with classical and quantum
subsystems by the joint tomographic probability distributions
depending on random classical and quantum positions and to propose
evolution equation for the tomograms of such system states
compatible with Liouville and von Neumann kinetic equations in
classical and quantum domains, respectively.

The paper is organized as follows. In Sec.2 the property of a joint
probability distributions \\$w(X_1,X_2,\theta_1,\theta_2)$ of two
random variables $ X_1$ and $X_2$ depending on two extra real
parameters $\theta_1$ and $\theta_2$ and their relation to optical
tomograms of a system with two particles are studied. In Sec.3 the
joint tomographic probability for a system with classical and
quantum subsystems is introduced. In Sec.4 the evolution equation
for the tomogram of hybrid system with classical and quantum
subsystems is proposed on the base of evolution equation for  the
optical tomograms of the separated classical and quantum particles
found in \cite{Korennoy,AmosovKorennoy}. In Sec.5 conclusions and
perspectives are discussed and entanglement properties of hybrid
systems are shortly considered.

\section{Correlations of random variables}
In this Section we study optical tomographic joint probability
distribution depending on two random variables. We associate the
probability distribution with state of system containing two
subsystems. Given a joint probability distributions
$w(X_1,X_2,\theta_1,\theta_2)$ of two real random variables $X_1$
and $X_2$ and two angles $\theta_1$,$\theta_2$. The random variables
are positions of two particles measured in reference frame in phase
space with rotated axes, rotation angles being labeled by $\theta_1$
and $\theta_2$. The joint probability distribution called optical
tomogram satisfies the following conditions of positivity
\begin{eqnarray}\label{eq.1.1}
w(X_1,X_2,\theta_1,\theta_2)\geq 0,
\end{eqnarray}
and normalization
\begin{eqnarray}\label{eq.1.2}
\int w(X_1,X_2,\theta_1,\theta_2)dX_1dX_2=1.
\end{eqnarray}
There exist following possibilities. If the two particles are
classical ones the joint probability distributions describing a
state of the particle obeys the positivity condition of the
following integral which is the Radon integral transform connecting
the tomogram with the probability density describing the system
state
\begin{eqnarray}
&&\frac{1}{4\pi^2}\int
\limits_0^{\pi}\int\limits_0^{\pi}d\theta_1d\theta_2\int\limits^{+\infty}_{-\infty}d\eta_1d\eta_2dX_1dX_2w(X_1,X_2,\theta_1,\theta_2)|\eta_1\eta_2|
\nonumber\\
&&\times\exp{i[\eta_1(X_1-q_1\cos\theta_1-p_1\sin\theta_1)
+\eta_2(X_2-q_2\cos\theta_2-p_2\sin\theta_2)]}=f(q_1,p_1,q_2,p_2)\geq
0.\label{eq.1.3}
\end{eqnarray}
The function $f(q_1,p_1,q_2,p_2)$ is the probability density of two
particles on their phase-space. This function is normalized
\begin{eqnarray}
\int f(q_1,p_1,q_2,p_2)dq_1dq_2dp_1dp_2=1\label{eq.1.4}
\end{eqnarray}
if the conditional (\ref{eq.1.2}) holds. If the tomogram describes a
state of two quantum particles it must satisfy the positivity
condition for the density operator given by the following integral
which is quantum version of Radon transform
\begin{eqnarray}
&&\hat{\rho}(1,2)=\frac{1}{4\pi^2}\int
\limits_0^{\pi}\int\limits_0^{\pi}d\theta_1d\theta_2\int\limits^{+\infty}_{-\infty}d\eta_1d\eta_2dX_1dX_2w(X_1,X_2,\theta_1,\theta_2)|\eta_1\eta_2|
\nonumber\\
&&\times\exp{i[\eta_1(X_1-\hat{q_1}\cos\theta_1-\hat{p_1}\sin\theta_1)
+\eta_2(X_2-\hat{q_2}\cos\theta_2-\hat{p_2}\sin\theta_2)]}\geq
0.\label{eq.1.5}
\end{eqnarray}
Here $\hat q_1$,$\hat q_2$,$\hat p_1$,$\hat p_2$ are position and
momentum operators of both quantum particles, respectively. The
inequality (\ref{eq.1.5}) means that the eigenvalues of the operator
$\hat{\rho}(1,2)$ are nonnegative numbers. Using the tomographic
probability description of classical and quantum states by means of
joint probability distribution one can naturally consider the hybrid
situation. Let the tomogram $w(X_1,X_2,\theta_1,\theta_2)$ satisfy
the following conditions. The tomogram of first particle
$\Omega_1(X_1,\theta_1)=\int w(X_1,X_2,\theta_1,\theta_2)dX_2$ and
tomogram of second particle $\Omega_2(X_2,\theta_2)=\int
w(X_1,X_2,\theta_1,\theta_2)dX_1$ must satisfy the conditions
\begin{eqnarray}
\frac{1}{2\pi}\int\limits_0^{\pi}d\theta_1\int\limits^{+\infty}_{-\infty}d\eta_1dX_1\Omega_1(X_1,\theta_1)|\eta_1|
\exp{i\eta_1(X_1-q_1\cos\theta_1-p_1\sin\theta_1)}=f_1(q_1,p_1)\geq
0\label{eq.1.6}
\end{eqnarray}
and
\begin{eqnarray}
\hat{\rho}(2)=\frac{1}{2\pi}\int\limits_0^{\pi}d\theta_2\int\limits^{+\infty}_{-\infty}d\eta_2dX_2\Omega_2(X_2,\theta_2)|\eta_2|
\exp{[i\eta_2(X_2-\hat{q_2}\cos\theta_2-\hat{p_2}\sin\theta_2)]}\geq
0.\label{eq.1.7}
\end{eqnarray}
These conditions mean that the integral (\ref{eq.1.6}) provides the
probability density on phase-space of the classical first particle
and the integral (\ref{eq.1.7}) provides density operator of the
quantum state of the second particle. The joint tomogram can satisfy
(\ref{eq.1.3}) and violate (\ref{eq.1.5}). Such tomogram describes
classical states of two particles. Another possibility corresponds
to case where the tomogram satisfies (\ref{eq.1.5}) and violates
(\ref{eq.1.3}). Such tomogram describes quantum states of two
particles. The joint tomogram $w(X_1,X_2,\theta_1,\theta_2)$ can
describe neither classical nor quantum state of two particles
violating both conditions (\ref{eq.1.3}) and (\ref{eq.1.5}). The
tomograms could be used to describe hybrid system of two particles
one of which is classical and another one is quantum.

\section{Correlations of quantum and classical variables}

\noindent The tomogram of hybrid system can be written in factorized
form
\begin{eqnarray}\label{eq.1.8}
w(X_1,X_2,\theta_1,\theta_2)
=\Omega(X_1,\theta_1)\Omega_2(X_2,\theta_2).
\end{eqnarray}
Here $\Omega_1(X_1,\theta_1)$ is classical tomogram and
$\Omega_2(X_2,\theta_2)$ is quantum tomogram. The tomogram
(\ref{eq.1.8}) is joint probability distribution of two random
positions $X_1$ and $X_2$ of the system state which does not contain
correlations of these observables. It means that behaviour of
classical particle does not influence on the behaviour of the
quantum particle and vice versa. But there exists the tomograms of
the form
\begin{eqnarray}\label{eq.1.8a1}
w(X_1,X_1,\theta_1,\theta_2)=P\Omega_1(X_1,\theta_1)\Omega_2(X_2,\theta_2)+(1-P)\bar
\Omega_1(X_1,\theta_1)\bar\Omega_2(X_2,\theta_2),
\end{eqnarray}
where $0$$\leq$ $P$$\leq1$. The tomogram (\ref{eq.1.8}) describes
the state which is mixture (convex sum) of two joint probability
distribution (\ref{eq.1.5}) without correlations. The mixture
provides the nonzero correlation of two random positions. In fact
the covariance of two random positions reads
\begin{eqnarray}\label{eq.1.9}
&&\sigma_{X_1 X_2}=\langle X_1X_2\rangle-\langle X_1\rangle\langle
X_2\rangle= \nonumber\\ &&\int w(X_1,X_2,\theta_1,\theta_2)X_1X_2d
X_1d X_2-\int X_1 w(X_1,X_2,\theta_1,\theta_2)dX_1dX_2\int X_2
w(X_1,X_2,\theta_1,\theta_2)dX_1dX_2=\nonumber\\
&&\int(P\Omega_1(X_1,\theta_1)\Omega_2(X_2,\theta_2)+(1-P)\bar
\Omega_1(X_1,\theta_1)\bar\Omega_2(X_2,\theta_2))X_1X_2dX_1dX_2\nonumber\\
&&-\int X_1(P\Omega_1(X_1,\theta_1)\Omega_2(X_2,\theta_2)+(1-P)\bar
\Omega_1(X_1,\theta_1)\bar\Omega_2(X_2,\theta_2))dX_1dX_2\nonumber\\
&&\times\int
X_2(P\Omega_1(X_1,\theta_1)\Omega_2(X_2,\theta_2)+(1-P)\bar
\Omega_1(X_1,\theta_1)\bar\Omega_2(X_2,\theta_2))dX_1dX_2\label{eq.1.9}
\end{eqnarray} and it is not equal to zero.

General expression for the tomogram of the state of the hybrid
system, i.e. the generic convex sum of the tomograms without
correlations reads
\begin{eqnarray}\label{eq.1.10}
w(X_1,X_2,\theta_1,\theta_2) =\sum_k P_k
\Omega_1^{(k)}(X_1,\theta_1)\Omega_2^{(k)}(X_2,\theta_2)
\end{eqnarray}
where $0\leq P_k\leq1$ and $\sum_k P_k=1$. This tomogram corresponds
to the state with covariance
\begin{eqnarray}
&&\sigma_{X_1X_2}=\langle
X_1X_2\rangle-\langle X_1\rangle\langle X_2\rangle= \nonumber\\
&&\int w(X_1,X_2,\theta_1,\theta_2)X_1X_2dX_1dX_2-\int X_1
w(X_1,X_2,\theta_1,\theta_2)dX_1dX_2\int X_2
w(X_1,X_2,\theta_1,\theta_2)dX_1dX_2=\nonumber\\
&&\int(\sum
P_k\Omega_1^{(k)}(X_1,\theta_1)\Omega_2^{(k)}(X_2,\theta_2))X_1X_2dX_1dX_2\nonumber\\
&&-\int X_1(\sum
P_k\Omega_1^{(k)}(X_1,\theta_1,\Omega_2^{(k)}(X_2,\theta_2))dX_1dX_2\nonumber\\
&&\times\int X_2(\sum
P_k\Omega_1^{(k)}(X_1,\theta_1,\Omega_2^{(k)}(X_2,\theta_2))dX_1dX_2\label{eq.1.11}
\end{eqnarray}
The formula (\ref{eq.1.10}) can be generalized to the case of
entangled states of the hybrid system. The tomogram of the entangled
state by analogy of the entangled state of quantum bipartite system
reads
\begin{equation}\label{eq.1.10a}
w^{ent}(X_1,X_2,\theta_1,\theta_2)=(1+\mu)\sum_{k} P_{k}
\Omega_1^{(k)}(X_1,\theta_1)\Omega_2^{(k)}(X_2,\theta_2)-\mu\sum_{k'}
P_{k'} \Omega_1^{(k')}(X_1,\theta_1)\Omega_2^{(k')}(X_2,\theta_2)
\end{equation}
where $\mu\geq0$.

The entangled tomogram is analogous to the form of distribution
given by two numbers $z$ and $1-z$, $1\geq z\geq0$ which is obtained
as the difference \begin{eqnarray*}
&&z=(1+\mu)x-\mu y;\nonumber\\
&&1-z=(1+\mu)(1-x)-\mu(1-y),\nonumber\\
&& \quad 1\geq x,y\geq0\end{eqnarray*} The conditions
\[y(\frac{\mu}{1+\mu})\leq x\leq\frac{1+y\mu}{\mu+1}\]
guarantee that probability distribution determined by number $z$ is
similar to the probability distribution for entangled states.

\section{Evolution equation}
The tomogram of the hybrid system has to satisfy the evolution
equation. We suggest the following equation which contains two
contributions
\begin{eqnarray}
&& \frac{\partial}{\partial t}w(X_1,X_2,\theta_1,\theta_2,
t)=\nonumber\\
&&[\cos^2\theta_1\frac{\partial}{\partial\theta_1}-\frac{1}{2}\sin2\theta_1
\{1+X_1\frac{\partial}{\partial
X_1}\}]w(X_1,X_2,\theta_1,\theta_2)\nonumber\\
&&+2[\mbox{Im}U_1\{q_1\rightarrow\{\sin\theta_1\frac{\partial}{\partial\theta_1}[\frac{\partial}{\partial
X_1}]^{-1}
+X_1\cos\theta_1+i\frac{\sin\theta_1}{2}\frac{\partial}{\partial
X_1}\}\}]w(X_1,X_2,\theta_1,\theta_2)\nonumber\\
&&+[\cos^2\theta_2\frac{\partial}{\partial\theta_2}-\frac{1}{2}\sin2\theta_2
\{1+X_2\frac{\partial}{\partial
X_2}\}]w(X_1,X_2,\theta_1,\theta_2)\nonumber\\
&&+[\frac{\partial}{\partial q_2}U_2\{q_2\rightarrow
\{\sin\theta_2\frac{\partial}{\partial\theta_2}[\frac{\partial}{\partial
X_2}]^{-1}
+X_2\cos\theta_2+i\frac{\sin\theta_2}{2}\frac{\partial}{\partial
X_2}\}\}]\sin\theta_2\frac{\partial}{\partial X_2}
w(X_1,X_2,\theta_1,\theta_2).\label{eq.1.12}
\end{eqnarray}
Here $U_1$ and $U_2$ are potential energy for quantum and classical
particles respectively.

For symplectic tomogram of the two-particle state $w(X_1,
X_2,\mu_1,\mu_2,\nu_1,\nu_2,t)$ the evolution equation
(\ref{eq.1.12}) has the form
\begin{eqnarray}
&&\{\frac{\partial}{\partial t}
-\nu_1\frac{\partial}{\partial\mu_1}-\nu_2\frac{\partial}{\partial\mu_2}
-\frac{1}{i}[U_1\{q_1\rightarrow(-\frac{\partial}{\partial\mu_1}
(\frac{\partial}{\partial
X_1})^{-1}+\frac{i}{2}\nu_1\frac{\partial}{\partial X_1})\}-c.c.]\nonumber\\
&&-[\frac{\partial U_2}{\partial
q_2}\{q_2\rightarrow-\frac{\partial}{\partial\mu_2}(\frac{\partial}{\partial
X_2})^{-1}\}]\nu_2\frac{\partial}{\partial
X_2}\}w(X_1,X_2,\mu_1,\mu_2,\nu_1,\nu_2,t)=0,\label{eq.1.14}
\end{eqnarray}
Both equations (\ref{eq.1.12}) and (\ref{eq.1.14}) provide the
classical Liouville evolution for the classical particle tomogram
and quantum von Neumann evolution for quantum particle tomogram.
These two tomograms are obtained by averaging the tomogram of two
particles either with respect to quantum position variable $X_1$ or
classical position variable $X_2$.

The suggested equations preserve the form of tomograms given both by
convex sum of tomograms without correlations and by the entangled
state tomograms. On the other hand the correlation of classical and
quantum observables are present for solutions of the equations.

\section{Conclusions}
To conclude we summarize the main results of the paper. We suggested
to describe the states of a system containing a classical and a
quantum subsystems by means of the joint tomographic probability
distributions and studied the properties of such tomograms. We
proposed the evolution equation for the states of such hybrid
quantum-classical system. The evolution equation is shown to provide
the Liouville equation for the classical subsystem tomogram and von
Neumann evolution equation for quantum subsystem after corresponding
averaging in the proposed equation on the quantum and classical
degrees of freedom, respectively. We shown that the entanglement of
the classical and quantum subsystems of the hybrid system can be
formulated in terms of properties of the joint tomographic
probability of the bipartite classical-quantum system. The partial
case of Gaussian tomogram for two-mode electromagnetic field in the
states which are generalizations of quantum separable or entangled
states is worthy to study. One can check experimentally presence of
classical mode by means of studying homodyne quadrature distribution
which for classical mode can violate uncertainty relation. We point
out that the suggested formalism provides extension of conventional
quantum and classical mechanics.

\section{Acknowledgement}
V.N.C. and V.I.M. thank Russian Foundation for Basic Research for
the support under Project Nos.07-02-00598,08-02-90300, and
09-02-00142


\begin{thebibliography}{99}

\bibitem{Sch26} E.~Schr\"odinger, {\sl Ann.~Phys} (Leipzig), {\bf 79}, p.~489
(1926)

\bibitem{Landau27}L. D. Landau, {\sl Z. Physik}, {\bf 45}, p. 430 (1927)

\bibitem{vonNeumann27}J. von Neumann, {\it Mathematische Grundlagen der Quantenmechanik},
Springer, Berlin (1932)

\bibitem{ManciniPhysLet96}S. Mancini, V. I. Man'ko, and P. Tombesi, {\sl Phys. Lett. A},
{\bf 213}, 1 (1996)

\bibitem{Ibort2009}A. Ibort, V. I. Man'ko, G. Marmo, A. Simoni, and
F. Ventriglia, {\sl Phys. Scripta}, {\bf 79}, p. 065013 (2009)

\bibitem{Olga1997}O.~V.~Man'ko, V.~I.~Man'ko, {\sl J. Russ. Laser Research}, {\bf 18}, N 2,
407 (1997)

\bibitem{MendesVIMPhysicaD}V. I. Man'ko, R. V. Mendes, {\sl Physica D}, {\bf 145},
p. 330 (2000)

\bibitem{Raymer} D. T. Smithey, M. Beck, M. G. Raymer, A. Faridani, {\sl Phys. Rev. Lett.},
{\bf 70}, p. 1244 (1993)


\bibitem{LvovskiRaymer} A. I. Lvovsky and M. G. Raymer, {\sl Rev.
Mod. Phys.}, {\bf 81}, p. 299 (2009)

\bibitem{SolimenoPorzio} 
V. I. Man'ko, G. Marmo, A. Porzio, S. Solimeno, F. Ventriglia, {\sl
Physica Scripta}, {\bf 83} 045001 (2011)

\bibitem{Olga2004}O. V. Man'ko, V. I. Man'ko, {\sl J. Russ. Laser Research}, {\bf 25},
N 2, 477 (2004)


\bibitem{Peres199?}A. Peres, D. Terno, {\sl Phys. Rev. A}, {\bf 63}, 022101 (2001)



\bibitem{Elze1} H-T. Elze, G. Gambarotta, F. Vallone, {\sl J. Phys.: Conf.Ser.},
{\bf 306}, 012010 (2011)



\bibitem{Wig32} E.~Wigner, {\sl Phys.~Rev.}, {\bf 40}, 749
(1932)

\bibitem{Korennoy} Yu. A. Korennoy, V. I. Man'ko, "Probability
representation of quantum evolution and energy level equations for
optical tomograms" e-Print: arXiv:1101.2537v1[quant-ph] 13 Jan. 2011

\bibitem{AmosovKorennoy}
G. G. Amosov, Ya. A. Korennoy, V. I. Man'ko, "Operators and their
symbols in the optical probabilistic representation of quantum
mechanics" e-Print: arXiv:1104.5606[quant-ph] Apr 2011. 11 pp.



\end{thebibliography}
\end{document}